\documentclass[11pt,twoside]{article}
\usepackage{./asp2014}
\usepackage{graphicx}
\usepackage[usenames, dvipsnames]{color}

\aspSuppressVolSlug
\resetcounters

\begin{document}

\title{How Do Cold Gas Outflows Shape Galaxies?}
\author{Alberto D. Bolatto,$^1$ Lee Armus,$^2$ Adam K. Leroy,$^3$ Sylvain Veilleux,$^4$ Fabian Walter,$^5$ and Richard Mushotzky$^6$ \\
\affil{$^1$University of Maryland, Department of Astronomy, College Park, MD 20742, USA; \email{bolatto@astro.umd.edu}}
\affil{$^2$Infrared Processing and Analysis Center, California Institute of Technology, 1200 E. California Boulevard, Pasadena, CA 91125, USA; \email{lee@ipac.caltech.edu}}
\affil{$^3$Ohio State University, Department of Astronomy, Columbus, OH 43210, USA; \email{leroy.42@osu.edu}}
\affil{$^4$University of Maryland, Department of Astronomy, College Park, MD 20742, USA; \email{veilleux@astro.umd.edu}}
\affil{$^5$Max Planck Institute for Astronomy, Heidelberg, Germany; \email{walter@mpia.de}}
\affil{$^6$University of Maryland, Department of Astronomy, College Park, MD 20742, USA; \email{richard@astro.umd.edu}}
}
\paperauthor{Alberto Bolatto}{bolatto@umd.edu}{}{University of Maryland}{Department of Astronomy}{College Park}{MD}{20742}{USA}
\paperauthor{Adam Leroy}{leroy.42@osu.edu}{}{Ohio State University}{Department of Astronomy}{Columbus}{OH}{43210}{USA}
\paperauthor{Sample~Author3}{Author3Email@email.edu}{ORCID_Or_Blank}{Author3 Institution}{Author3 Department}{City}{State/Province}{Postal Code}{Country}


\section{The Importance of Galaxy-scale Winds}


It has become increasingly clear over the last two decades that it is impossible to understand how galaxies form and evolve through cosmic time without a deeper understanding of how 
star formation and black hole accretion affect the growth of galaxies and the state of their gas reservoirs. The failure of both large-scale cosmological simulations and detailed physical simulations of individual galaxies to reproduce the observed characteristics of the galaxy population has driven home the point - understanding how feedback works is probably the key open question in galaxy evolution. This is intimately linked to one of the fundamental questions identified in the {\em ``New Worlds, New Horizons''} decadal report: how do baryons cycle in
and out of galaxies? 

Galactic winds are thought to shape the galaxy mass function, heat the circumgalactic medium, play a critical role in quenching star formation, and pollute the intergalactic medium with heavy elements \citep[e.g.,][]{VEILLEUX2005}.
Galaxies are not closed systems, and winds are necessary to explain chemical evolution, as well as playing a key role in shaping the disk-halo interface. The fastest and most energetic winds arise due to feedback from active galactic nuclei (AGN) or strong starbursts, while less 
powerful, and more localized, galactic fountains due to clustered star formation recycle material between the disk and the halo even in less active galaxies. On smaller scales, super star clusters can also drive outflows that shape their final stellar mass and affect their environment \citep[e.g.,][]{HERRERA2017}. 

\begin{figure}[t]
\includegraphics[width=\textwidth]{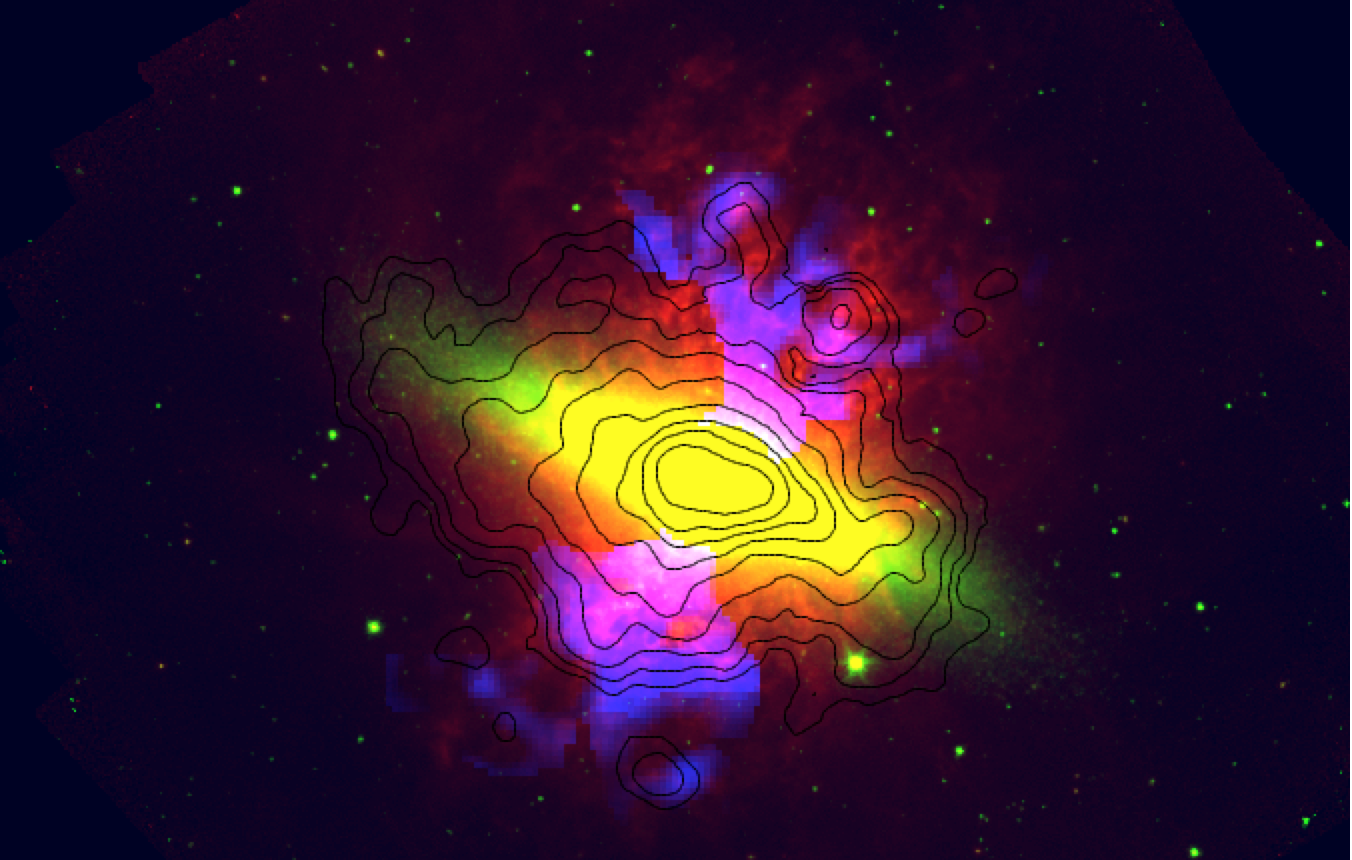}
\caption{The cooler phases of the starburst-driven wind in the prototypical starburst M\,82. The red, green, and blue colors correspond respectively to PAH emission (from IRAC 8.0 $\mu$m), the stellar component (from IRAC 3.6 $\mu$m), both from \citep{KENNICUTT2003}, and HI emission kinematically identified to be part of the outflow \citep[from][]{LEROY2015} imaged at 24\arcsec \ resolution. The black contours show the CO emission (in this case CO $2-1$) at 20\arcsec\ resolution obtained by the IRAM 30m telescope. The contours correspond to both the disk and the outflow emission \citep[also from][]{LEROY2015}. \label{fig:M82fig}}
\end{figure}

Detailed studies of nearby systems have frequently focused on the warm or hot phases of galactic winds, which are often visible in X-rays or optical emission lines. 
{The manifestation of these hot winds can often be quite spectacular, consisting of huge, bipolar nebulae \citep[e.g.,][]{STRICKLAND2004}, but they typically carry very little mass $M_{Xray}\sim10^6$\,M$_\odot$ \citep[e.g.,][]{LEHNERT1999}. One of the important effects of winds is to shut down star formation by removal of gas, but this hot gas is very low density compared to the cold gas out of which young stars will ultimately form.} But galactic winds are multi-phase phenomena (Figure \ref{fig:M82fig}).
When present, colder phases, containing neutral atomic and dense molecular gas, can dominate the mass budget of the outflow \citep{WALTER2002,RUPKE2005,FERUGLIO2010,ALATALO2011,RUPKE2013}. 

Until recently, observations of the cooler phases of galactic outflows have been hindered by a lack of sensitivity and/or spatial resolution to properly image the low surface brightness wind and unambiguously connect it to the processes in the disk that power the outflow.  With Herschel, it has been possible to detect fast, dense outflows in a number of local Ultraluminous Infrared and starburst galaxies \citep[e.g.,][]{STURM2011,VEILLEUX2013}.  In some cases the mass outflow rates are comparable to, or larger than, the star formation rates of the galaxies themselves, suggesting a significant impact on the lifetime of the active phase. However, with Herschel it was only possible to detect the nearest, brightest sources, and the winds were nearly always unresolved. ALMA and the IRAM Plateau de Bure Interferometer are 
currently making important advances in finding and imaging molecular outflows \citep{BOLATTO2013a,COMBES2013,CICONE2014,SAKAMOTO2014,GARCIA-BURILLO2014,ZSCHAECHNER2016,VEILLEUX2017}, but the resolution and surface brightness sensitivity requirements make identifying and characterizing molecular winds challenging with present-day facilities. 

\section{The Need for the ngVLA}

The ngVLA will have the sensitivity and frequency coverage to enable a giant leap in our understanding of the cold phases of galactic winds. The ngVLA will be able to: 1) 
identify molecular and neutral atomic winds using sensitive, high resolution imaging, 2) characterize their mass, outflow 
rate, and physical state using molecular multi-line spectroscopy and 21\,cm emission, 3) diagnose their fate by imaging their extent with high surface brightness 
sensitivity across multiple phases and providing detailed kinematics, and 4) study their physical launching mechanisms. 

Imaging galactic outflows requires the combination of high spectral sensitivity, resolution, and the ability to recover a wide range of spatial scales -- all things for which the ngVLA excels. Close to the source, typical column densities may correspond to optical extinction ($A_V$) of a few magnitudes, rapidly decaying with distance from the source as the outflow expands and gets diluted. The hydrogen column density corresponding to $A_V\sim 1$ is ${\rm N(H)}=1.9\times10^{21}$\,cm$^{-1}$ \citep{BOHLIN1978,RACHFORD2009}. At $\lambda=21$\,cm the ngVLA is capable of attaining an atomic hydrogen column density RMS of ${\rm N(H)}\sim5.2\times10^{20}$\,cm$^{-1}$ in 10\,km\,s$^{-1}$ in one hour, at a resolution of 1\arcsec. Therefore assuming a linewidth of 30 km/s and an integration time of 24 hours, it can detect and image atomic outflows down to ${\rm N(H)}\sim3\times10^{20}$\,cm$^{-1}$ at $5\sigma$, which corresponds to $A_V\sim0.1$ mag in 30\,km\,s$^{-1}$ (equivalent to a surface density of $\Sigma_{\rm HI}\sim1.5$\,M$_\odot$\,pc$^{-2}$, or a mass ${\rm M(HI)}\sim2.5\times10^4$\,M$_\odot$ at D=30\,Mpc). 

The molecular content of outflows may decrease faster than their atomic content, as molecular cloudlets get shattered and dissociated \citep[see, for example,][]{LEROY2015}. At $\lambda=3$~mm the ngVLA has a line sensitivity of 10 mK in 10 km/s at a resolution of 1\arcsec. Assuming a CO $1-0$ emission with linewidth of 30 km/s, and a CO-to-H$_2$ factor of $X_{\rm CO}\sim0.5\times10^{20}$ cm$^{-2}$\,(K\,km\,s$^{-1}$)$^{-1}$  \citep[intermediate between Milky Way GMCs and optically thin emission,][]{BOLATTO2013a,BOLATTO2013b}, the ngVLA will be capable of detecting molecular outflows down to $A_V\sim0.01$ mag at $5\sigma$ (equivalent to $\Sigma_{\rm H2}\sim0.15$\,M$_\odot$\,pc$^{-2}$, or ${\rm M(H_2)}\sim2.5\times10^3$\,M$_\odot$ at D=30 Mpc) in 24 hours at 1\arcsec\ resolution. 
ALMA is capable of comparable 
Rayleigh-Jeans brightness temperature sensitivities in CO $J=3-2$.  However, the factor of $\sim27$ higher critical density required to excite the $J=3-2$ transition compared to $J=1-0$ means that only very dense molecular gas will be bright in the CO $J=3-2$ line. As we emphasize below, the combination of the ngVLA and ALMA provides a powerful set of tools for studying a wide range of density and excitation conditions in molecular outflows.

Some of the key open questions that ngVLA observations will help answer are: 
\begin{enumerate}
\item What are the driving mechanisms of cold winds? Mechanical feedback from supernovae \citep[e.g.,][]{FUJITA2009}, radiation pressure \citep[e.g.,][]{MURRAY2011}, cosmic ray pressure gradients \citep[e.g.,][]{UHLIG2012}, entrainment facilitated by Kelvin-Helmholtz instabilities \citep[e.g.,][]{HECKMAN2000}, and direct driving by interaction with AGN jets \citep[e.g.,][]{WAGNER2011} have all been proposed as ways to inject momentum in the gas. It remains unclear, however, how they combine, and which one if any dominates \citep{HOPKINS2012,MURATOV2015}. A combination of high-resolution and sensitivity observations of the cold phases as they are ejected are key to solve this problem. For example, it appears that radiation pressure is insufficient to explain the high-resolution properties of the NGC~253 molecular outflow \citep{WALTER2017}.

\item What is the fate of the launched gas? Is molecular gas reformed in the wind? Does expelled gas change phases? 
Imparting momentum to molecular cloudlets without destroying them has proven difficult in numerical simulations, while at the same time it is expected that part of the hot phase outflow may cool and reform a cold phase \citep[e.g.,][]{THOMPSON2016}. Observations of the outflow in M~82 strongly suggest that there is conversion of molecular into neutral atomic gas as the outflowing gas progresses away from the galaxy \citep{LEROY2015}. High sensitivity, resolved imaging in neutral atomic gas and molecular tracers is needed to answer these questions.

\item What are the wind mass loss rates? What is the 
mass-loss to star-formation-rate ratio (the mass loading parameter)? Are mass-loading parameters and mass-loss rates as high as predicted for low-mass galaxies? These parameters are key inputs to cosmological simulations, necessary to understand the precise effects of feedback on galaxy growth. Current estimates of the efficiency with which momentum is imparted to the different gas phases, implemented as sub-grid recipes in physical galaxy simulations, suggest mass loading parameters of order $\eta\sim10$ for massive galaxies, and as high as $\eta\sim100$ for dwarf galaxies \citep{MURATOV2015}. While higher mass galaxies {are expected to} retain most of their metals in their circum-galactic environment, dwarf galaxies should {heavily pollute} 
the IGM \citep{MURATOV2017}. Such values of $\eta$ appear to be necessary to reproduce galaxy properties in cosmological simulations. It remains unclear, however, how to precisely attain such high mass loading efficiencies in detailed simulations \cite{KIM2018}. Most of the mass loss is due to the denser, cold phases, likely dominated by the neutral atomic and molecular gas that is directly observable with the ngVLA. {The key to measuring accurate molecular mass loss rates is an understanding of the density structure in the wind, critically dependent on observations of several molecular species in the outflowing gas}.  These observations require extremely high sensitivity, wide bandwidth, and the ability to spatially and kinematically map the emission from the outflow.

\item What are the conditions triggering cool outflows? 
Observations suggest the existence of a star formation surface density threshold for launching large outflows \citep[e.g.,][]{NEWMAN2012}, and a similar threshold in luminosity appears to exist for AGN-launched outflows \citep[e.g.,][]{VEILLEUX2013,WYLEZALEK2017}. Lower velocity galactic fountains and even radiation pressure-driven outflows, can also occur over extended areas of disks. Systematic demographics of the different properties of the multiphase outflows and their hosts, collected through a combination of radio and other multi-wavelength observations {(see below), will provide valuable information} about the conditions and triggers of outflows.
 
\item What fraction of the {outflowing gas} reaches escape velocity versus falling back and being reaccreted/recycled? Detailed simulations suggest that winds and fountains go through phases, dominated alternatively by outflow and inflow \citep{KIM2018}. Observations show velocity gradients in the expelled gas, that can be interpreted in some cases as acceleration \citep{WALTER2017}, or deceleration \citep{MARTINI2018}. Generally, recycled material may play an important role in lengthening the gas depletion timescale of galaxies \citep{DAVE2011}, feeding galaxies at late cosmic times, and {allowing for the exchange of} processed gas between galaxies \citep{ANGLES2017b}. {Understanding the fraction of escaping gas and the relative amounts of expelled and recycled material, 
requires high sensitivity and high angular resolution observations of molecular and atomic gas tracers in galaxies and their circum-galactic environments.}

\item {Are winds only effective at suppressing the formation of stars (negative feedback), or can they also trigger star formation in galaxies?} Although outflows driven by star formation or AGN are frequently invoked as mechanisms to expel gas and/or quench star formation \citep[e.g.,][]{ALATALO2015}, star formation can also be enhanced by compressive turbulence driven by the mechanical energy input of the wind \citep[e.g.,][]{VANBREUGEL1985,CROFT2006}. 
These positive feedback processes can also have an important effect on galaxy evolution \citep{SILK2013}. 
Multi-wavelength radio techniques are particularly well suited to studying this problem, since they can be used to simultaneously {image the outflowing gas and measure the gas kinematics, as well as derive the star formation rate (via the thermal free-free radio continuum emission) on sub-kpc scales.}

\item How do winds {affect} 
the growth of black holes? Simulations suggest that black hole growth, particularly at early times, is limited by stellar feedback, which 
expels gas from 
galactic nuclei, {limiting accretion}. As a consequence, black holes can be undermassive in low-mass galaxies with respect to their high-mass counterparts, causing them to {fall below} the M$_{BH}-\sigma_{halo}$ relation \citep[e.g.,][]{ANGLES2017a}. {The ability of the ngVLA to resolve and measure gas content and kinematics in the innermost regions of galaxies, 
is key to quantifying the effects of winds on the growth of supermassive black-holes.} 

\item What 
are the statistical properties of cold winds in the universe? 
Are these a rare phenomenon 
confined to AGN and starbursts, or are they a general feature of galaxies? What is the
redshift evolution of starburst- and quasar-driven winds?  
Observations suggest that galactic winds are ubiquitous at high redshift \citep[e.g.,][]{NEWMAN2012}. We know very little about their cold components, except in a handful of spectacular examples \citep[e.g.,][]{MAIOLINO2012}. Fast outflows are seen in powerful IR galaxies in the local Universe \citep{VEILLEUX2013}, but studies have been limited to a handful of the brightest, most energetic sources. Large, sensitive, systematic surveys of molecular outflows {reaching out to epochs where most galaxies are rapidly evolving are necessary to establish the importance of outflows in a cosmological context}.  

\item
What is the relation between the large scale cold outflows and
the AGN that seem to trigger them? Are the ultra-fast outflows (UFOs)
seen in the X-rays on sub-parsec scale the ultimate drivers of the
most powerful cold outflows, or are the more common but slower soft
X-ray and ultraviolet (UV) warm absorbers and UV broad absorption line
(BAL) outflows seen on larger scale than the UFOs a better predictor
of these cold outflows? How is the energy in these nuclear winds
transferred to drive the galaxy-scale cold outflows?
The kinetic power {that is available from an AGN wind} 
scales with $v_{out}^3$, rising quite rapidly with outflow velocity.
{Thus the fastest outflows are the ones that can potentially produce the largest feedback effects, provided they couple with the ISM effectively}. X-ray observations identify a type of outflow that can reach modestly relativistic speeds $v_{out}\sim0.03\,c-0.3\,c$, consisting of a highly ionized flow originating from the accretion disk itself with column densities as high as $10^{24}$\,cm$^{-2}$ \citep{REEVES2003,TOMBESI2010}. {In theory,} these very fast accretion-disk winds can drive shocks into the host galaxy ISM and create shock-driven over-pressurized bubbles that give rise to the large-scale outflows observed in ionized, neutral, and molecular gas \citep{FAUCHER-GIGUERE2012,TOMBESI2015}. 
Testing these models, and directly linking fast accretion disk outflows with galactic winds  
requires high-quality, velocity-resolved imaging of the molecular and neutral atomic gas at high spatial resolution in the central regions of AGN with identified X-ray UFOs.
A more complete survey of AGN will help relate the cold outflows to the more prevalent warm absorbers \citep{CRENSHAW2012} and Broad Absorption Line (BAL) outflows  \citep{GIBSON2009}. The prospects are good that ngVLA and future UV-optical facilities will be able to spatially resolve the regions where the BAL outflows interact with the ISM of the AGN host galaxies \citep{MOE2009,BAUTISTA2010,DUNN2010}.

\end{enumerate}

\section{The Unique Discovery Space of the ngVLA}
The ngVLA has a unique part of the discovery space to fill, of surface brightness sensitivity at high enough 
resolution to spatially resolve outflows close in and trace them far out from galaxy centers while at the 
same time studying the launching mechanisms and compositions.

Obtaining these measurements requires imaging and kinematics of the cool molecular and atomic component. This requires high surface brightness 
sensitivity on physical scales of $20-100$\,pc out to 100 Mpc, with good imaging dynamic range (frequently 
the faint wind emission in next to bright line or continuum emission from the galaxy/AGN). A distance of 100\,Mpc 
encloses a volume big enough to find several examples of LIRGs/ULIRGs and powerful AGN. Physical resolution of $20-100$\,pc ($\sim20$ pc is the width of the streamers in NGC253) is necessary to study the base of the outflows and their launching mechanisms, 
while high surface brightness sensitivity allows us to follow the outflow further out and search for accelerations/velocity 
gradients that constrain the final velocity. Even higher resolution is useful to study the disk/nucleus next to the 
outflow (usually much brighter), and investigate the connections to energy sources such as super star clusters, collections of supernova remnants, or SMBH accretion disks. To find and trace the extent 
and composition of winds it is necessary to image the CO, HI, and OH lines along with the continuum, as well as a host of tracers of local physical conditions. To characterize the physical 
state of winds and their mass loss, CO isotopologues, shock tracers, probes of the diverse excitation mechanisms (IR/UV/CR/X-ray) and tracers of 
the density distribution are key \citep[e.g.,][]{LINDBERG2016,WALTER2017}.

\section{Complementarity with Existing and Planned Facilities}

Three of the NASA flagship mission concepts under study have significant complementarity for the study of galaxy outflows, which are naturally multi-phase, multi-wavelength phenomena. In the following we focus on the complementary capabilities brought by the Infrared {\em Origins Space Telescope} (OST), the X-ray Lynx Telescope, and the Large UV/Optical/Infrared Surveyor (LUVOIR), while also discussing the {\em Square Kilometer Array} (SKA) and the Atacama Large Millimeter/submillimeter Array (ALMA). 

{With a huge leap in line and surface brightness sensitivity over all previous and planned IR missions, high spatial resolution, and the ability to quickly map large areas of the sky to
generate unbiased spectroscopic samples, the OST will be able to find cool outflows spanning from the present day to over 90\% of the age of the Universe.  {These targets will be prime targets for detailed follow-up study with the ngVLA at high angular and spectral resolution.} For example, the Medium Resolution Survey Spectrometer (MRSS) on OST can detect columns of extra-planar neutral gas with $N_{HI}\sim 10^{20}$\,cm$^{-2}$ and $n_H \sim 1$\,cm$^{-3}$ via [CII] at sub-kpc resolution out to 30 Mpc in 10 minutes, and fine-structure lines in fast moving galactic winds down to $\sim 2x10^{-21}$\,W\,m$^{-2}$ in one hour, {mapping nearby galaxies and detecting sources well into the epoch of re-ionization.} For example, outflow signatures such as P-Cygni profiles (blueshifted absorption and redshifted emission) are easily detectable in the far-infrared molecular OH feature at $\lambda_{rest} =79\,\mu$m from a source like Mrk~231 out to $z\sim5$ in an hour, making surveys of large numbers of galaxies feasible in the far-infrared for the first time.  Through blind, low-resolution spectroscopic surveys and detailed studies of nearby starbursts and AGN, OST working together with the ngVLA, will deliver a complete picture of feedback on the cold and warm atomic and molecular ISM in rapidly evolving galaxies. }
{The proposed JAXA/ESA mission SPICA, recently selected as one of three candidate missions to be further studied for ESA's M5 mission call, will also be an extremely capable observatory for studying galactic outflows in the infrared \citep{GONZALEZ-ALFONSO2017}, although it will have significantly less collecting area than OST and a narrower wavelength coverage.}

While OST searches for cool outflows, Lynx will characterize the highly ionized $10^6-10^7$~K plasma phase (although measurements in absorption can also measure cold phases). The crucial capabilities of the Lynx telescope for galactic winds are high angular resolution ($\sim0.5\arcsec$), high spectral resolution of $\sim2$ eV FWHM independent of energy (corresponding to $R\sim700$ at the very strong OVII line, and high signal-to-noise imaging over a field of view of $\sim5\arcmin$. Lynx also has a grating with energy-resolution of 0.3 eV, capable of resolving the kinematics of the hot wind phases, and producing density diagnostics for the soft X-ray emitting winds. This combination of parameters allows measurement of dynamics (velocities of $\sim200$\,km\,s$^{-1}$), chemical composition (through lines from C, N, O, Ne, Mg, Si, S, Fe, and Ni), and diagnostics of ionization state. This allows Lynx to study the narrow emission-line regions of hot plasma bubbles inflated by AGN, as well as measuring metallicities outside the cores of massive clusters out to a redshift $z\sim3$, and determining the mechanisms of gas ionization. In dusty winds Lynx can also distinguish dust features in absorption as the binding energy of atoms in grains is $\sim1$\,eV, allowing it to detect whether oxygen is bound in grains or in the gas phase. Lynx's planned sensitivity will allow the study of M~82-like starbursts at moderate redshifts, the most luminous starbursts at $z\sim2-3$, and lensed objects at $z\sim4$. In addition Lynx's high sensitivity and resolution in the Fe K band will allow a detailed analysis of the ultrafast outflows, determining their energetics and geometry over a wide range of redshifts, luminosities and black hole masses.  Similarly, ESA's Advanced Telescope for High Energy Astrophysics (ATHENA) to be launched in the early 2030s, will also be able to detect and measure the hot gas associated with galactic outflows. However, ATHENA's significantly lower spatial and spectral resolution compared to Lynx (5\arcsec\ and 2.5\,eV compared to 1\arcsec\ and 0.3\,eV, respectively) will make it better suited to studying outflows from powerful AGN, and likely limit its usefulness for low-velocity, starburst-driven winds.


LUVOIR will bring a leap in sensitive high-resolution imaging and spectroscopy across the full UVOIR (0.1 $-$ 3 $\mu$m) bandpass, with the capability to  
resolve structures on scales of 100 parsecs or smaller
in all galaxies at $z \le 4$.  The excellent match in resolution
between LUVOIR and ngVLA will allow for direct comparisons between the
warm and cool phases of the outflows and their impact on these
galaxies. In particular, LUVOIR will provide spatially resolved maps
of the UV BAL outflows in quasars to directly probe the mechanisms
involved in driving the most extreme cold outflows.  The 50--100 times larger UV
spectroscopic sensitivity of LUVOIR compared with {\em HST} will bring
a hundredfold more UV-bright quasars within its reach. It will also be able to use the much more numerous UV-bright
galaxies as background sources.  With this dramatic gain in
sensitivity, LUVOIR will be able to efficiently probe
gas and metals outside of galaxies, and the complex interplay between accretion flows onto galaxies and gas outflowing from the galaxies. These results will nicely complement the low surface brightness maps of galaxies and their halos produced by ngVLA.

The mid-frequency SKA has significant overlap with the low-frequency capabilities of the ngVLA, and so it will be able to perform HI and low-frequency OH studies of outflowing gas. The SKA phase 1 is comparable to the ngVLA, with a somewhat smaller collecting area but without the high-frequency capabilities. The precise sensitivity comparison will depend on decisions about array configurations, and the upper frequency cutoff of the SKA. However, it is clear that much of the molecular and high-resolution science associated with the outflows will be out of reach of even a phase 2 mid-frequency SKA. With a full mid-frequency SKA built, the complementarity with the ngVLA will be excellent, with the former focusing on the low frequency science while the ngVLA observes at higher frequencies ($\nu>10-20$~GHz).

Compared to the the ngVLA, ALMA, which currently 
{drives much of the research on high-resolution studies of molecular outflows} in galaxies, will have significantly lower line sensitivity in the region of frequency overlap (ALMA bands 1 to 3). But because of its extensive high frequency coverage the combination of ALMA and ngVLA will be extremely complementary, particularly for studies of and launching mechanisms and chemistry and excitation in outflows. 

\acknowledgements ADB acknowledges support from NSF-AST 1412419. 


\begin{thebibliography}{}
\bibitem[Alatalo et al.(2011)]{ALATALO2011} Alatalo, K., Blitz, L., Young, L.~M., et al.\ 2011, \apj, 735, 88 
\bibitem[Alatalo et al.(2015)]{ALATALO2015} Alatalo, K., Lacy, M., Lanz, L., et al.\ 2015, \apj, 798, 31 
\bibitem[Angl{\'e}s-Alc{\'a}zar et al.(2017a)]{ANGLES2017a} Angl{\'e}s-Alc{\'a}zar, D., Faucher-Gigu{\`e}re, C.-A., Quataert, E., et al.\ 2017, \mnras, 472, L109 
\bibitem[Angl{\'e}s-Alc{\'a}zar et al.(2017b)]{ANGLES2017b} Angl{\'e}s-Alc{\'a}zar, D., Faucher-Gigu{\`e}re, C.-A., Kere{\v s}, D., et al.\ 2017, \mnras, 470, 4698 
\bibitem[Bautista et al.(2010)]{BAUTISTA2010} Bautista, M. A., Dunn, J. P., Arav, N., Korista, K. T., Moe, M., \& Benn, C.\ 2010, \apj,713, 553
\bibitem[Bohlin et al.(1978)]{BOHLIN1978} Bohlin, R.~C., Savage, B.~D., \& Drake, J.~F.\ 1978, \apj, 224, 132
\bibitem[Bolatto et al.(2013a)]{BOLATTO2013a} Bolatto, A.~D., Wolfire, M., \& Leroy, A.~K.\ 2013, \araa, 51, 207 
\bibitem[Bolatto et al.(2013b)]{BOLATTO2013b} Bolatto, A.~D., Warren, S.~R., Leroy, A.~K., et al.\ 2013, \nat, 499, 450
\bibitem[van Breugel et al.(1985)]{VANBREUGEL1985} van Breugel, W., Filippenko, A.~V., Heckman, T., \& Miley, G.\ 1985, \apj, 293, 83 
\bibitem[Cicone et al.(2014)]{CICONE2014} Cicone, C., Maiolino, R., Sturm, E., et al.\ 2014, \aap, 562, A21 
\bibitem[Combes et al.(2013)]{COMBES2013} Combes, F., Garc{\'{\i}}a-Burillo, S., Casasola, V., et al.\ 2013, \aap, 558, A124 
\bibitem[Crenshaw \& Kraemer(2012)]{CRENSHAW2012} Crenshaw, D. M., \& Kraemer, S. B.\ 2012, \apj, 753, 75
\bibitem[Croft et al.(2006)]{CROFT2006} Croft, S., van Breugel, W., de Vries, W., et al.\ 2006, \apj, 647, 1040
\bibitem[Dav{\'e} et al.(2011)]{DAVE2011} Dav{\'e}, R., Oppenheimer, B.~D., \& Finlator, K.\ 2011, \mnras, 415, 11
\bibitem[Dunn et al.(2010)]{DUNN2010} Dunn, J. P., Bautista, M., Arav, N., et al.\ 2010, \apj, 709, 611
\bibitem[Faucher-Gigu{\`e}re \& Quataert(2012)]{FAUCHER-GIGUERE2012} Faucher-Gigu{\`e}re, C.-A., \& Quataert, E.\ 2012, \mnras, 425, 605 
\bibitem[Feruglio et al.(2010)]{FERUGLIO2010} Feruglio, C., Maiolino, R., Piconcelli, E., et al.\ 2010, \aap, 518, L155 
\bibitem[Fujita et al.(2009)]{FUJITA2009} Fujita, A., Martin, C.~L., Mac Low, M.-M., New, K.~C.~B., \& Weaver, R.\ 2009, \apj, 698, 693 
\bibitem[Garc{\'{\i}}a-Burillo et al.(2014)]{GARCIA-BURILLO2014} Garc{\'{\i}}a-Burillo, S., Combes, F., Usero, A., et al.\ 2014, \aap, 567, A125 
\bibitem[Gibson et al.(2009)]{GIBSON2009} Gibson, R. R., Jiang, L., Brandt, W. N., et al.\ 2009, \apj, 692, 758
\bibitem[Gonz{\'a}lez-Alfonso et al.(2017)]{GONZALEZ-ALFONSO2017} Gonz{\'a}lez-Alfonso, E., Armus, L., Carrera, F.~J., et al.\ 2017, PASA, 34, e054 
\bibitem[Heckman et al.(2000)]{HECKMAN2000} Heckman, T.~M., Lehnert, M.~D., Strickland, D.~K., \& Armus, L.\ 2000, \apjs, 129, 493 
\bibitem[Herrera \& Boulanger(2017)]{HERRERA2017} Herrera, C.~N., \& Boulanger, F.\ 2017, \aap, 600, A139 
\bibitem[Hopkins et al.(2012)]{HOPKINS2012} Hopkins, P.~F., Quataert, E., \& Murray, N.\ 2012, \mnras, 421, 3522 
\bibitem[Kennicutt et al.(2003)]{KENNICUTT2003} Kennicutt, R.~C., Jr., Armus, L., Bendo, G., et al.\ 2003, \pasp, 115, 928 
\bibitem[Kim \& Ostriker(2018)]{KIM2018} Kim, C.-G., \& Ostriker, E.~C.\ 2018, \apj, 853, 173 
\bibitem[Lehnert et al.(1999)]{LEHNERT1999} Lehnert, M.~D., Heckman, T.~M., \& Weaver, K.~A.\ 1999, \apj, 523, 575 
\bibitem[Leroy et al.(2015)]{LEROY2015} Leroy, A.~K., Walter, F., Martini, P., et al.\ 2015, \apj, 814, 83 
\bibitem[Lindberg et al.(2016)]{LINDBERG2016} Lindberg, J.~E., Aalto, S., Muller, S., et al.\ 2016, \aap, 587, A15 
\bibitem[Maiolino et al.(2012)]{MAIOLINO2012} Maiolino, R., Gallerani, S., Neri, R., et al.\ 2012, \mnras, 425, L66 
\bibitem[Martini et al.(2018)]{MARTINI2018} Martini, P., Leroy, A.~K., Mangum, J.~G., et al.\ 2018, arXiv:1802.04359 
\bibitem[Moe et al.(2009)]{MOE2009} Moe, M., Arav, N., Bautista, M. A., \& Korista, K. T.\ 2009, \apj, 706, 525
\bibitem[Muratov et al.(2015)]{MURATOV2015} Muratov, A.~L., Kere{\v s}, D., Faucher-Gigu{\`e}re, C.-A., et al.\ 2015, \mnras, 454, 2691 
\bibitem[Muratov et al.(2017)]{MURATOV2017} Muratov, A.~L., Kere{\v s}, D., Faucher-Gigu{\`e}re, C.-A., et al.\ 2017, \mnras, 468, 4170 
\bibitem[Murray et al.(2011)]{MURRAY2011} Murray, N., M{\'e}nard, B., \& Thompson, T.~A.\ 2011, \apj, 735, 66 
\bibitem[Newman et al.(2012)]{NEWMAN2012} Newman, S.~F., Genzel, R., F{\"o}rster-Schreiber, N.~M., et al.\ 2012, \apj, 761, 43
\bibitem[Rachford et al.(2009)]{RACHFORD2009} Rachford, B.~L., Snow, T.~P., Destree, J.~D., et al.\ 2009, \apjs, 180, 125 
\bibitem[Reeves et al.(2003)]{REEVES2003} Reeves, J.~N., O'Brien, P.~T., \& Ward, M.~J.\ 2003, \apjl, 593, L65 
\bibitem[Rupke \& Veilleux(2013)]{RUPKE2013} Rupke, D.~S.~N., \& Veilleux, S.\ 2013, \apj, 768, 75 
\bibitem[Rupke et al.(2005)]{RUPKE2005} Rupke, D.~S., Veilleux, S., \& Sanders, D.~B.\ 2005, \apjs, 160, 115 
\bibitem[Sakamoto et al.(2014)]{SAKAMOTO2014} Sakamoto, K., Aalto, S., Combes, F., Evans, A., \& Peck, A.\ 2014, \apj, 797, 90
\bibitem[Silk(2013)]{SILK2013} Silk, J.\ 2013, \apj, 772, 112 
\bibitem[Sturm et al.(2011)]{STURM2011} Sturm, E., Gonz{\'a}lez-Alfonso, E., Veilleux, S., et al.\ 2011, \apjl, 733, L16 
\bibitem[Strickland et al.(2004)]{STRICKLAND2004} Strickland, D.~K., Heckman, T.~M., Colbert, E.~J.~M., Hoopes, C.~G., \& Weaver, K.~A.\ 2004, \apjs, 151, 193 
\bibitem[Thompson et al.(2016)]{THOMPSON2016} Thompson, T.~A., Quataert, E., Zhang, D., \& Weinberg, D.~H.\ 2016, \mnras, 455, 1830 
\bibitem[Tombesi et al.(2010)]{TOMBESI2010} Tombesi, F., Cappi, M., Reeves, J.~N., et al.\ 2010, \aap, 521, A57 
\bibitem[Tombesi et al.(2015)]{TOMBESI2015} Tombesi, F., Mel{\'e}ndez, M., Veilleux, S., et al.\ 2015, \nat, 519, 436 
\bibitem[Uhlig et al.(2012)]{UHLIG2012} Uhlig, M., Pfrommer, C., Sharma, M., et al.\ 2012, \mnras, 423, 2374 
\bibitem[Veilleux et al.(2005)]{VEILLEUX2005} Veilleux, S., Cecil, G., \& Bland-Hawthorn, J.\ 2005, \araa, 43, 769 
\bibitem[Veilleux et al.(2013)]{VEILLEUX2013} Veilleux, S., Mel{\'e}ndez, M., Sturm, E., et al.\ 2013, \apj, 776, 27 
\bibitem[Veilleux et al.(2017)]{VEILLEUX2017} Veilleux, S., Bolatto, A., Tombesi, F., et al.\ 2017, \apj, 843, 18 
\bibitem[Wagner \& Bicknell(2011)]{WAGNER2011} Wagner, A.~Y., \& Bicknell, G.~V.\ 2011, \apj, 728, 29 
\bibitem[Walter et al.(2002)]{WALTER2002} Walter, F., Weiss, A., \& Scoville, N.\ 2002, \apjl, 580, L21 
\bibitem[Walter et al.(2017)]{WALTER2017} Walter, F., Bolatto, A.~D., Leroy, A.~K., et al.\ 2017, \apj, 835, 265 
\bibitem[Wylezalek et al.(2017)]{WYLEZALEK2017} Wylezalek, D., Schnorr M{\"u}ller, A., Zakamska, N.~L., et al.\ 2017, \mnras, 467, 2612 
\bibitem[Zschaechner et al.(2016)]{ZSCHAECHNER2016} Zschaechner, L.~K., Walter, F., Bolatto, A., et al.\ 2016, \apj, 832, 142 


\end{thebibliography}


\end{document}